%Paper: hep-th/9312176
%From: Hugh Osborn <H.Osborn@damtp.cambridge.ac.uk>
%Date: Wed, 22 Dec 93 09:25 GMT

%Plain TeX file

\magnification\magstep1
%\nopagenumbers
%For smaller font footnotes
\newdimen\itemindent \itemindent=32pt
\def\textindent#1{\parindent=\itemindent\let\par=\resetpar%
\indent\llap{#1\enspace}\ignorespaces}

\let\oldpar=\par
\def\resetpar{\oldpar\parindent=20pt\let\par=\oldpar}

\font\ninerm=cmr9 \font\ninesy=cmsy9
\font\eightrm=cmr8 \font\sixrm=cmr6
\font\eighti=cmmi8 \font\sixi=cmmi6
\font\eightsy=cmsy8 \font\sixsy=cmsy6
\font\eightbf=cmbx8 \font\sixbf=cmbx6
\font\eightit=cmti8
\def\eightpoint{\def\rm{\fam0\eightrm}
  \textfont0=\eightrm \scriptfont0=\sixrm \scriptscriptfont0=\fiverm
  \textfont1=\eighti  \scriptfont1=\sixi  \scriptscriptfont1=\fivei
  \textfont2=\eightsy \scriptfont2=\sixsy \scriptscriptfont2=\fivesy
  \textfont3=\tenex   \scriptfont3=\tenex \scriptscriptfont3=\tenex
  \textfont\itfam=\eightit  \def\it{\fam\itfam\eightit}%
  \textfont\bffam=\eightbf  \scriptfont\bffam=\sixbf
  \scriptscriptfont\bffam=\fivebf  \def\bf{\fam\bffam\eightbf}%
  \normalbaselineskip=9pt
  \setbox\strutbox=\hbox{\vrule height7pt depth2pt width0pt}%
  \let\big=\eightbig  \normalbaselines\rm}
\catcode`@=11 %
\def\eightbig#1{{\hbox{$\textfont0=\ninerm\textfont2=\ninesy
  \left#1\vbox to6.5pt{}\right.\n@space$}}}
\def\vfootnote#1{\insert\footins\bgroup\eightpoint
  \interlinepenalty=\interfootnotelinepenalty
  \splittopskip=\ht\strutbox %
  \splitmaxdepth=\dp\strutbox %
  \leftskip=0pt \rightskip=0pt \spaceskip=0pt \xspaceskip=0pt
  \textindent{#1}\footstrut\futurelet\next\fo@t}
\catcode`@=12 %

\hsize 6 true in
\vsize 8.5 true in
\baselineskip=14pt
\input mssymb.tex

\def \l{\langle}
\def \r{\rangle}
\def \vep{\varepsilon}

\def \ep{\epsilon}

\def \E{{\cal {E}}}
\def \I{{\cal {I}}}
\def \O{{\cal {O}}}
\def \T{{{\cal T}}}
\def \L{{{\cal L}}}
\def \R{{{\cal R}}}
\def \bO{\bar{\cal {O}}}
\def \hh{{1\over 2}}
\def \nab{\nabla}
\def \pr{\partial}
\def \ts{\textstyle}

\def \half{{\ts{1\over 2}}}
\def \si{\sigma}
\def \de{\delta}

\parindent 20pt
\parskip 8pt
\rightline{DAMTP 93/67}
\vskip 28pt
\centerline {\bf Implications of Conformal Invariance for Quantum Field
Theories in $d>2$}
\vskip 24pt
\centerline {H. OSBORN}
\vskip 5pt
\centerline {\it {DAMTP, University of Cambridge, Silver St.,}}
\centerline {\it {Cambridge CB3 9EW, England.}}
\vskip 20pt
\centerline {ABSTRACT}
%\vskip 2pt
\font \abs=cmr10 at 10 true pt
\font \absit=cmti10 at 10 true pt
{\abs \openup -1\jot %\baselineskip 8pt
{\narrower\smallskip\parindent 0pt
Recently obtained results for two and three point functions for
quasi-primary operators in conformally invariant theories in arbitrary
dimensions {\absit d} are described. As a consequence the three point function
for the energy momentum tensor has three linearly independent forms for
general {\absit d} compatible with conformal invariance. The corresponding
coefficients may be regarded as possible generalisations of the Virasoro
central charge to {\absit d} larger than 2. Ward identities which link two
linear combinations of the coefficients to terms appearing in the energy
momentum tensor trace anomaly on curved space are discussed.
The requirement of positivity for expectation values of the energy density is
also shown to lead to positivity conditions which are simple for a particular
choice of the three coefficients. Renormalisation
group like equations which express the constraints of broken conformal
invariance for quantum field theories away from critical points are
postulated and applied to two point functions.\hfill\break
Talk presented at the  XXVII Ahrenshoop International Symposium.

\narrower}
\openup 1\jot}

\vskip 11pt

Conformal  field theories in two dimensions have proved to be a remarkably
fruitful area of investigation in recent years, motivated initially by
applications to string theory but also of interest for systems in
statistical physics  at their critical points. In two dimensions there are
a very large class of explicitly constructed non trivial conformal field
theories which have a rich mathematical structure. In some sense in two
dimensions conformal field theories appear to be dense in the space of all
quantum field theories, under perturbation by relevant operators then there
usually appears to be a `nearby' conformal field theory defining an infra-red
fixed point of the renormalisation group beta functions and to which the
theory flows when the length scale is taken to infinity [1]. According to the
Zamolodchikov $c$-theorem [2] the value of the Virasoro central charge $c$
decreases from its initial value to that for the conformal field theory
defined by the critical point describing the massless fields in the large
distance limit, at least for unitary theories. The value of $c$ represents
a measure of the degrees of freedom in the conformal theory and its decrease
corresponds intuitively to a kind of `coarse-graining' in the limit of large
distance scales [3].

Such a detailed description of conformal field theories does not extend
to higher dimensions. Nevertheless for field theories in which the trace
of the energy momentum tensor $T_{\mu\nu}$ may be expanded in a basis of
scalar operators $\O_i$, $T_{\mu\mu} = \beta^i \O_i$, where $\beta^i$ are
the beta functions of the various couplings, then at a fixed point defined
by vanishing beta functions the trace of the energy momentum tensor is zero
and the theory enjoys conformal invariance.  The conserved charges have
the form $Q_v = \int dS_\mu \, T_{\mu\nu} v_\nu$ where $\pr_\mu v_\nu +
\pr_\nu v_\mu \propto \de_{\mu\nu} $ so that, as described later,
$v_\mu$ corresponds to an
infinitesimal conformal transformation. In general scale invariance
does not imply conformal invariance but for most quantum field theories
it may be expected to hold at any renormalisation group fixed point [4].
However for $d>2$ the detailed knowledge of non trivial conformal field
theories is much less explicit. Non gaussian fixed points may be investigated
in the $\vep$ expansion for $d=4-\vep$ and $d=2+\vep$ and also in the $1/N$
expansion for some models when $2<d<4$ [5]. In three dimensions conformal field
theories should correspond to the different universality classes related to the
possible critical points of realistic statistical
physical systems. In four dimensions conformal field theories are apparently
rather sparse, based on the triviality of the continuum limit of lattice
quantum field theories, although possible exceptions should be given by
$N=4$ supersymmetric gauge theories and also by large $N$ QCD, with gauge
group $SU(N)$, where the number of fermions in the fundamental representation
is fine tuned to ensure the leading beta function coefficient is negative
but ${\rm O}(1)$ (there is then a perturbatively accessible infra-red fixed
point with $g_*^2 N = {\rm O}(N^{-1})$).

With the motivation of exploring further the consequences of conformal
invariance in $d>2$ A. Petkou and I [6] have attempted to establish
a basis for further work by deriving simple conformally invariant forms
for the two and three point functions for
arbitrary spin operators and in particular apply these to the energy
momentum tensor which plays an essential role in any quantum field theory.
Although the conformal group is finite dimensional, being isomorphic to
$O(d+1,1)$ in the Euclidean case, there are significant complications as
compared with $d=2$ due to the basic representations of the spin group $O(d)$
acting on the quantum fields being no longer one dimensional.

The group of conformal transformations is defined by coordinate transformations
such that
$$ x_\mu \to x'{}_{\! \mu} (x) \, ,
\quad dx'{}_{\! \mu} dx'{}_{\! \mu} = \Omega(x)^{-2} dx_\mu dx_\mu \ .
\eqno (1) $$
For an infinitesimal transformation we may write
$$ \eqalign {
x'{}_{\! \mu} (x) = x_\mu + v_\mu (x) , \qquad \Omega (x) = 1 -
\si_v (x) \, , \cr
\quad \pr_\mu v_\nu + \pr_\nu v_\mu =  2 \si_v
\de_{\mu\nu} \, , \qquad \si_v = {1\over d}\, \pr {\cdot v} \, . \cr}
\eqno (2) $$
Except for $d=2$ the general solution of (2) has the form
$$ v_\mu (x) = a_\mu + \omega_{\mu\nu} x_\nu + \lambda x_\mu +
b_\mu x^2 - 2 x_\mu b {\cdot x} \, , \quad  \omega_{\mu\nu} = -
 \omega_{\nu\mu} \, , \quad \si_v(x) = \lambda - 2 b {\cdot x} \, ,
\eqno (3) $$
representing infinitesimal translations, rotations, scale transformations
and special conformal transformations. The corresponding Lie algebra is
easily determined from
$$ \eqalign {
[ v_1 , v_2 ]_\mu = {}& v_1 {\cdot \pr} \, v_{2\mu} -
v_2 {\cdot \pr} \, v_{1\mu} = v'{}_{\! \mu} \, , \cr
a'{}_{\! \mu} = {} & (\omega_2 a_1 - \omega_1 a_2 )_\mu + \lambda_2 a_{1\mu}
- \lambda_1 a_{2\mu} \, , \cr
\omega'{}_{\! \mu\nu} = {} & [ \omega_2 , \omega_1 ]_{\mu\nu}
+ 2 ( b_{2\mu} a_{1\nu} +  a_{2\mu} b_{1\nu} -  b_{1\mu} a_{2\nu}
- a_{1\mu} b_{2\nu} ) \, , \cr
\lambda' = {}& 2( a_2 {\cdot b_1} - a_1 {\cdot b_2} ) \, , \cr
b'{}_{\! \mu} = {} & (\omega_2 b_1 - \omega_1 b_2 )_\mu - \lambda_2 a_{1\mu}
+ \lambda_1 a_{2\mu} \, . \cr}
\eqno (4) $$
Besides conformal transformations connected to the identity it is important
to consider also inversions where
$$ x'{}_{\!\mu} = {x_\mu \over x^2} \, ,
\quad \Omega (x) = x^2 \ .
\eqno (5) $$
The whole conformal group may be generated just by combining inversions
with translations and rotations.
Geometrically the conformal group maps circles
to circles (or straight lines) so any three distinct points may be moved
to any other three points.

The quantum fields are assumed to transform under conformal transformations
as a simple extension of their transformations under translations and
rotations where
$$ \O^i(x) \to \O'{}^i (x') = D^i {}_{\! j} ( R ) \O^j (x) \, , \quad
x' = R x + a \, , \quad R^T R = 1 \, ,
\eqno (6) $$
with $D(R)$ belonging to some finite dimensional representation of $O(d)$.
For a general
conformal transformation as in (1) we may define an $O(d)$ rotation
depending on $x$ by
$$\R_{\mu \nu}(x) = \Omega (x)
{\pr x'{}_{\! \mu} \over \pr x_\nu} \ , \quad \R^T (x)
\R (x) =  1 \, ,
\eqno (7) $$
and then we take for quasi-primary fields  the transformation rule [7]
$$ \O^i(x) \to \O'{}^i (x') = \Omega(x)^{\eta}  D^i {}_{\! j}(\R (x) )
\O^j (x) \, ,
\eqno (8) $$
where $\eta$ is the scale dimension of the field. Infinitesimally (8)
is equivalent to
$$ \de \O (x) = - T_v (x) \O (x) \, , \quad T_v(x) = v_\mu (x)
{\pr \over \pr x_\mu} + \eta \, \si_v(x) + \half s_{\mu\nu}
\omega_{v\mu\nu} (x) \, ,
\eqno (9) $$
for $ \omega_{v\mu\nu} = - \half ( \pr_\mu v_\nu - \pr_\nu v_\mu ) $ and
where $s_{\mu\nu}$ are the generators of $O(d)$ in the appropriate
representation $([s_{\mu\nu},
s_{\si\rho}] = \de_{\mu\si} s_{\nu\rho} - \de_{\mu\rho} s_{\nu\si} +
\de_{\nu\rho} s_{\mu\si} - \de_{\nu\si} s_{\mu\rho})$ so that
$\R = 1 - \half s_{\mu\nu}\omega_{v\mu\nu}$. It is straightforward
to verify that $[T_{v_1} , T_{v_2} ] = T_{[v_1 , v_2]}$.

To give a general expression for two point functions for quasi-primary
fields it is convenient for each field $\O^i (x)$ to define a conjugate
field $\bO_i (x)$ transforming as
$$ \bO_i(x) \to \bO'{}_{\! i} (x') = \Omega(x)^{\eta}
\bO_j (x)  D^j {}_{\! i}(\R (x) ) \, .
\eqno (10) $$
Then if the representation of $O(d)$ to which $\O , \bO$ belong is
irreducible we may write
$$ \l \O^i (x) \, \bO_j (y) \r = {C_\O \over (s^2)^\eta} \,
D^i {}_{\! j}( I (s) ) \, , \quad s=x-y \, ,
\eqno (11) $$
where $C_\O$ is an overall constant scale factor and $I_{\mu\nu}(x)$
is the matrix belonging to $O(d)$ defined according to (7) corresponding
to the inversion (5),
$$ I_{\mu \nu}(x) = \de_{\mu \nu} - 2 \, {x_\mu x_\nu \over x^2} \, .
\eqno (12) $$
To verify (11) it is sufficient, since translations and rotations are trivial,
to use
$$ I_{\mu \alpha} (x) I_{\alpha \beta} (s ) I_{\beta \nu}
(y) = I_{\mu \nu} (s') \, , \quad s' = {x \over x^2}
- {y \over y^2} \ .
\eqno (13) $$
As an illustration [8] we consider a vector operator $V_\mu$ and the traceless
energy momentum tensor $T_{\mu\nu}$ where application of (11) gives
$$ \eqalign {
\l V_{\mu} (x) \, V_\nu (y) \r = {}&{C_V\over s^{2(d-1)}} \, I_{\mu \nu}
(s) \, , \quad
\l T_{\mu \nu}(x) \, T_{\si \rho} (y)\r = {C_T\over s^{2d}} \,
\I_{\mu \nu,\si \rho} (s) \, , \cr
\I_{\mu \nu,\si \rho} & (s) = \half \bigl ( I_{\mu \si}(s) I_{\nu \rho}
(s) + I_{\mu \rho}(s) I_{\nu \si} (s) \bigl ){} - {1\over d}\, \de_{\mu
\nu} \de_{\si \rho} \ , \cr}
\eqno (14) $$
where the dimensions of $V_\mu$ and $T_{\mu\nu}$ are respectively $d-1$
and $d$, as is necessary for conservation. For $n_S$ free scalar fields
and $n_D$ free Dirac fermions with arbitrary $d$
$$ S_d^{\, 2} C_T = n_S {d\over d-1} + n_F \half d N_d \, ,
\eqno (15) $$
where $S_d = 2\pi^{\hh d}/\Gamma(\hh d)$ and $N_d$ is the dimension of the
Dirac matrices in dimension $d$ while in four dimensions $\pi^4 C_T =
{1\over 3} n_S + 2n_F + 4n_V$ for $n_V$ free vector fields.

The conformally covariant form for three point functions is also
relatively simple when written in terms of an analogous group theoretic
expression [6,9],
$$ \eqalign {
\l \O_1^{i}(x)& \, \O_2^{j} (y) \, \O_3^{k} (z) \r  \cr
& = {1\over (x-z)^{2\eta_1}\,(y-z)^{2\eta_2}} \,
D_1^{\, i} {}_{i'} (I(x-z))
D_2^{\, j} {}_{j'} (I(y-z)) \, t_{12,3}^{i'j'k} (X) \, , \cr }
\eqno (16) $$
where
$$ X = {x-z \over (x-z)^2 } - {y-z \over (y-z)^2 } \, , \quad
X^2 = {(x-y)^2 \over (x-z)^2 (y-z)^2 }
\eqno (17) $$
and $t_{12,3}^{ijk} (X)$ is a homogeneous group invariant function
satisfying
$$ \eqalign {
D_1^{\, i} {}_{i'} (R) D_2^{\, j} {}_{j'} (R)
D_3^{\, k} {}_{ k'} (R) & \, t_{12,3}^{i'j'k'} (X) =
t_{12,3}^{ijk}(RX) \ \hbox {for all}\ R \in O(d) \ ,\cr
t_{12,3}^{ijk}(\lambda X) = {}& \lambda^{\eta_3 - \eta_1 - \eta_2}\,
t_{12,3}^{ijk}(X) \, . \cr}
\eqno (18) $$
To see how the required conformal transformation properties of (16)
follow from (18) it is important to
note that $X$, as defined in (17), transforms as a vector, $X_\mu \to
X'{}_{\! \mu} = \Omega(z) \R_{\mu\nu} (z) X_\nu$ and also $ D^i {}_{i'}
(\R(x)) D^{i'} {}_{j'} (I(x-z)) D^{-1 \, j'} {}_{\! j} (\R(z)) =
D^{i} {}_{\! j} (I(x'-z'))$. The representation displayed in (16) is not
manifestly symmetric but different permutations may be obtained in a
similar form by using
$$ t_{21,3}^{jik}(X) = t_{12,3}^{ijk}(-X) \, , \quad
t_{13,2}^{ikj} (X) = (X^2)^{\eta_2 - \eta_3} D_1^{\, i} {}_{i'} (I(X))\,
t_{12,3}^{i'jk}(X)\, .
\eqno (19) $$

It is not difficult to see that (16) gives for the leading term in
the short distance operator product expansion
$$ \O_1^i (x) \, \O_2^j (y) \sim
{1\over C_{\O_3}} t_{12,3}^{ijk}(s)\,
\bO_{3\, k}^{\vphantom g} (y) \, , \quad s=x-y \, ,
\eqno (20) $$
for $s\to 0$. Conversely using conformal invariance we may derive (16)
directly from (20) [10]. If we consider an inversion through a point $w$,
so that $x' = (x-w)/(x-w)^2$, $x^{\prime 2} = {1/(x-w)^2}$ etc, and then
let $w\to z$ so that $z'\to \infty$ we get
$$ \eqalign { \!\!\!\!\!\!
\l  \O_1^{i}& (x)\, \O_2^{j} (y) \, \O_3^{k} (z) \r \cr  ={}&
x^{\prime \, 2\eta_1} y^{\prime \, 2\eta_2} z^{\prime \, 2\eta_3}
\, D_1^{\, i} {}_{ i'} (I(x')) D_2^{\, j} {}_{ j'} (I(y'))
D_3^{\, k} {}_{\! k'} (I(z'))
\l \O_1^{i'}(x')\, \O_2^{j'} (y') \, \O_3^{k'} (z') \r \cr
\sim {}&
x^{\prime \, 2\eta_1} y^{\prime \, 2\eta_2} z^{\prime \, 2\eta_3}
D_1^{\, i} {}_{ i'} (I(x')) D_2^{\, j} {}_{ j'} (I(y')) \,
{1\over C_{\O_3}} t_{12,3}^{i'j'\ell}(s')\,
D_3^{\, k} {}_{\! k'} (I(z')) \l \bO_{3\, \ell}(y')
\O_3^{k'} (z') \r , \cr}
\eqno (21) $$
which in the limit gives just (16) using the form (11) for the two point
function
$ D_3^{\, k} {}_{\! k'} (I(z')) \l \bO_{3\, \ell}(y')\, \O_3^{k'} (z') \r
\sim  C_{\O_3}\de^k{}_{\!\ell}/ z^{\prime 2\eta_3}$ and noting that $s' =
x' - y' \to X$.

Additional constraints arise from the imposition of conditions such as
conservation equations. Applying (16) to the case when one of the
operators is a current $V_\mu$ of dimension $d-1$ so that we may write
$$ \eqalign {
\l V_\mu(x)& \, \O_2^{j} (y) \, \O_3^{k} (z) \r  \cr
& = {1\over (x-z)^{2(d-1)}\,(y-z)^{2\eta_2}} \,
I_{\mu\nu}(x-z) \, D_2^{\, j} {}_{j'} (I(y-z)) \, t_{\nu}{}^{j'k} (X) \, , \cr}
\eqno (22) $$
then $\pr_\mu V_\mu = 0$ requires
$$ \pr_\mu t_{\mu}{}^{jk} (X) = 0 \, .
\eqno (23) $$
Similar conditions arise from $\pr_\mu T_{\mu\nu} = 0$.

Using this formalism it is easy to recover, for instance, old results for
the conformally invariant form for three point functions involving three
vector currents [11]. As a new application we considered determining the
possible conformally invariant forms for the three point function for the
energy momentum tensor. Applying (12) gives
$$ \eqalign { \!\!\!\!\!\!\!\!
\l T_{\mu \nu}&(x) \,T_{\si\rho} (y) \, T_{\alpha \beta}  (z) \r \cr
& = {1\over (x-z)^{2d}\,(y-z)^{2d}} \,
\I_{\mu\nu,\mu'\nu'}(x-z) \, \I_{\si\rho,\si'\rho'}(x-y) \,
t_{\mu'\nu'\si'\rho'\alpha \beta} (X) \, , \cr}
\eqno (24) $$
where $t_{\mu\nu\si\rho\alpha \beta} (X)$ is symmetric and traceless on
each pair of indices $\mu\nu$, $\si\rho$ and $\alpha\beta$, and is also
symmetric under $\mu\nu \leftrightarrow \si\rho$ and is homogeneous of degree
$-d$. There are 8 independent forms, compatible with transforming
covariantly under $O(d)$ rotations, satisfying these conditions while
the conservation equation
$$ \pr_\mu t_{\mu\nu\si\rho\alpha \beta} (X) = 0
\eqno (25) $$
provides 5 independent relations. It is also necessary to impose the
condition
$$ \I_{\mu\nu,\mu'\nu'}(X)t_{\mu'\nu'\si\rho\alpha\beta}(X)=
t_{\alpha\beta\mu\nu\si\rho}(X) \,
\eqno (26) $$
to ensure that the expression (24) is completely symmetric. However the
relations so obtained are equivalent to those obtained from the conservation
condition (25) so there remain in general 3 linearly independent forms
for the conformally invariant energy momentum tensor three point function.

There is no obvious natural basis for the three independent forms but it is
possible to specify the general form for
$\l T_{\mu \nu}(x) \,T_{\si\rho} (y) \, T_{\alpha \beta}  (z) \r$
in terms of particular components for
some convenient choice of $x,y,z$. The conformal invariant expressions
become particular simple when $x,y,z$ lie on a straight line (it is
possible to use the collinear configuration as a starting point for a
general analysis). In consequence if we let $x=0, \, z={\bar y}$, where
${\bar y}$ denotes the reflection of $y$ through the plane $x_1 =0$,
then for $y_i = 0, \, i=2,3,\dots \ y_1 = - {\bar y}_1 = \half s$, we may take
$$ \eqalign { \!\!\!\!\!\!\!\!\!\!\!\!\!\!\!\!\!\!\!\!
\l T_{i1}(y) \,T_{11} (0) \, T_{k1} ({\bar y}) \r = {}&
{2^{2d} \over s^{3d}} \, \gamma \, \de_{ik} \, , \cr
\!\!\!\!\!\!\!\!\!\!\! \l T_{ij}(y) \,T_{11} (0) \, T_{k\ell} ({\bar y}) \r
= {}& {2^{2d} \over s^{3d}}  \biggl \{
\ep \Bigl ( \de_{ik}\de_{j\ell} + \de_{i\ell} \de_{jk} - {2\over d-1}
\de_{ij}\de_{k\ell} \Bigl ){} - {1\over d-1} \beta \, \de_{ij}\de_{k\ell}
\biggl \}  , \cr}
\eqno (27) $$
where $\beta,\gamma,\epsilon$ are parameters which determine the general
expression or equivalently the form of $t_{\mu\nu\si\rho\alpha \beta} (X)$.
For $d=2$ $\ep$ is irrelevant and $\beta=\gamma$ while for $d=3$ there is
the restriction $\gamma =
-4\epsilon$ so that respectively there are only one, two linearly independent
conformally invariant forms in these dimensions.

For free field theories these coefficients can be routinely calculated. For
general $d$ scalars and fermions give
$$ \eqalignno {
\!\!\!\!\!\!\! S_d^{\, 3} \beta = {}& n_S \, {1\over 8(d-1)^3} \bigl (
8d + 4 (d^2-1)(d-2)^2 + d(d-1)^3(d-2) + 5d(d^2-1)(d-2) \bigl)  \cr
& ~~~~~~~~~ + n_F\, {1\over 4}(d-1)(d+2)N_d \, \cr
\!\!\!\!\!\!\! S_d^{\, 3} \gamma ={}& n_S \, {d\over 8(d-1)^2} \bigl ( 4d(d-1)
+
(d+1)(d-2)^2 \bigl ){} + n_F \, {d\over 16}\bigl( 2d + (d-1)(d+2) \bigl )
N_d \, , \cr
\!\!\!\!\!\!\!\!\!\!
- S_d^{\, 3} \ep = {}& n_S \, {d\over 8(d-1)^3} \bigl ( 4d +(d-1)(d-2)^2 \bigl
)
{} +  n_F\, {d\over 8} (d-1)N_d \, , & (28) \cr}
$$
while when $d=4$ and $N_4=4$
$$ \eqalign {
S_4^{\, 3} \beta = {}& {\ts {136\over 27}} n_S + 18n_F + 16n_V \, , \cr
S_4^{\, 3} \gamma ={}& {\ts {34\over 9}} n_S + 26n_F + 32n_V \, , \cr
- S_4^{\, 3} \ep = {}& {\ts {14\over 27}} n_S + 6n_F + 32n_V \, . \cr}
\eqno (29) $$
Clearly these free field theories realise the full range of possibilities
in this case [12]. Restrictions are obtained by using Ward identities. These
may be derived by extending the quantum field theory to curved space with
metric $g_{\mu\nu}$ and postulating
$$ \nab^\mu \l T_{\mu\nu}\r = 0 \, , \quad g^{\mu\nu} \l T_{\mu\nu} \r =0 \, .
\eqno (30) $$
Taking functional derivatives gives relations between $n$ and $n-1$ point
functions, for example
$$\eqalign {
0 =  - {2\over \sqrt{g(y)}} {\de\over\de g^{\si\rho}(y)} \nab^\mu
\l T_{\mu\nu}\r = {}& \nab^\mu \l T_{\mu \nu}(x) \, T_{\si \rho} (y)\r
- \pr_\nu \de^d(x,y)\, \l T_{\si\rho} (x)\r \cr
& -\bigl \{ \nab_\si \bigl ( \de^d(x,y) \l T_{\rho\nu} (x)
\r \bigl ){} + \si \leftrightarrow \rho \bigl\} \, , \cr
0=  - {2\over \sqrt{g(y)}} {\de\over\de g^{\si\rho}(y)} g^{\mu\nu}
\l T_{\mu\nu} \r = {}& g^{\mu\nu} \l T_{\mu \nu}(x) \, T_{\si \rho} (y)\r
- 2\de^d(x,y) \l T_{\si\rho} (x) \r \, . \cr}
\eqno (31) $$
Restricting to flat space, since then $\l T_{\mu\nu} \r =0$, gives
$$ \pr_\mu \l T_{\mu \nu}(x) \, T_{\si \rho} (y)\r = 0 \, , \quad
\l T_{\mu \mu}(x) \, T_{\si \rho} (y)\r = 0 \, ,
\eqno (32) $$
and also we may obtain less trivial identities relating two
and three point functions. The additional terms in the Ward identities
for the three point function involving $\pr_\mu T_{\mu \nu}(x)$ and
$ T_{\mu \mu}(x)$ arise from the singularities in
$\l T_{\mu \nu}(x) \,T_{\si\rho} (y) \, T_{\alpha \beta}  (z) \r$
when $x\to y$ and $x\to z$. From (24) it is necessary and sufficient to
require
$$ \eqalign { \!\!\!\!\!\!\!\!
\pr_\mu t_{\mu\nu\si\rho\alpha\beta} (s) = {}& C_T \bigl (
\E_{\si\rho,\alpha\beta} \, \pr_\nu + \E_{\nu\rho,\alpha\beta}\, \pr_\si
+ \E_{\nu\si,\alpha\beta}\, \pr_\rho \bigl ) \de^d (s) \, , \cr
\!\!\!\!\!\!\!\! t_{\mu\mu\si\rho\alpha\beta} (s) = {}& 2C_T
\E_{\si\rho,\alpha\beta} \, \de^d (s) \, , \quad \E_{\mu\nu,\si\rho}
= \half \bigl (\de_{\mu \si} \de_{\nu \rho} + \de_{\mu \rho}
\de_{\nu \si} \bigl){} - {1\over d} \, \de_{\mu \nu} \de_{\si \rho} \, .  \cr}
\eqno (33) $$
To verify (33) requires careful regularisation of
$t_{\mu\nu\si\rho\alpha\beta} (s)$  so that it is a well defined distribution.
It is convenient to use the ideas of differential regularisation of
extracting derivatives which reduces the degree of the singularity as
$s\to 0$ so that it is integrable on ${\Bbb R}^d$ [13]. The resulting
distribution is arbitrary up to terms $\propto \de^d(s)$ but these are fixed
by imposing (33) which then entails
$$ C_T = {2S_d\over d(d-1)(d+2)} \bigl ( d\beta + 2(d-1) \gamma
- (d-2)(d+1) \ep \bigl ) \, .
\eqno (34) $$

As is well known [14] in two and four dimensions there are contributions
to the trace of the energy momentum tensor depending on the spatial
curvature even for field theories which are conformal on flat space.
For four dimensional conformal theories
$$ \eqalign {
g^{\mu\nu}\l T_{\mu\nu} \r = {}&
- \beta_a \, C_{\alpha\beta\gamma\de} C^{\alpha\beta\gamma\de} -
\beta_b \, {\ts {1\over 4}} \ep^{\mu\nu}{}_{\! \si\rho}\ep^{\alpha\beta}
{}_{\! \gamma\de} R_{\mu\nu\alpha\beta} R^{\si\rho\gamma\de} \, , \cr
C_{\alpha\beta\gamma\de} & = R_{\alpha\beta\gamma\de} - \half \bigl (
g_{\alpha[\gamma}R_{\delta]\beta} - g_{\beta[\gamma}R_{\delta]\alpha} \bigl )
{} + {\ts{1\over 3}} g_{\alpha[\gamma}g_{\delta]\beta}  R \, . \cr}
\eqno (35) $$
In this case there are then extra purely local terms in the trace identity
linking the two and three point functions
$$ \eqalign { \!\!\!\!\!\!\!\!\!\!\!\!
\l T_{\mu\mu} (x)  \, T_{\si\rho} (y) \, T_{\alpha \beta} (z) \r &{}
= 2  \bigl (\de^4(x-y)+\de^4(x-z)\bigl)\,
\l T_{\si\rho} (y) \, T_{\alpha \beta} (z) \r \cr
& \!\!\!\!\! \, - 32 \beta_a\,
\E^C{}_{\!\! \si\kappa\lambda\rho,\alpha\gamma\de\beta}\, \pr_{\kappa}
\pr_{\lambda} \de^4(x-y) \pr_{\gamma} \pr_{\de} \de^4(x-z) \cr
&  \!\!\!\!\! \, + 4 \beta_b \bigl \{
\ep_{\si\alpha\gamma\kappa}\ep_{\rho\beta\delta\lambda}\pr_\kappa
\pr_\lambda \bigl( \pr_\gamma \de^4 (x-y) \pr_\de \de^4 (x-z) \bigl ) {}
+ \si \leftrightarrow \rho \bigl \} ,\cr }
\eqno (36) $$
where $\E^C{}_{\!\! \mu\si\rho\nu,\alpha\beta\gamma\delta}$ is the
projection operator onto 4 index tensors with the symmetries of the
Weyl tensor $C_{ \mu\si\rho\nu}$ ($ C_{\mu\si\rho\nu} =
C_{[\mu\si][\rho\nu]}, \, C_{\mu[\si\rho\nu]} = 0, \, C_{\mu\si\rho\mu} = 0$).
{}From (36) and the usual renormalisation group equation it is easy to derive
$$ \eqalign {
\mu{\pr \over \pr \mu} \l T_{\si\rho} (y) \, T_{\alpha \beta} (z) \r = {}&
\int \! d^4x \, \l T_{\mu\mu} (x)\, T_{\si\rho} (y) \, T_{\alpha \beta} (z) \r
- 4 \l T_{\si\rho} (y) \, T_{\alpha \beta} (z) \r \cr
= {}& -32\beta_a \,\E^C{}_{\!\! \si\kappa\lambda\rho,\alpha\gamma\de\beta}\,
\pr_{\kappa} \pr_{\lambda} \pr_{\gamma} \pr_{\de} \de^4 (y-z) \, , \cr}
\eqno (37) $$
which relates $\beta_a$ to the scale of the two point function
$$ \beta_a = - {\pi^2 \over 640} \, C_T \, .
\eqno (38) $$
Clearly (36) provides an additional condition on the energy momentum
tensor three point function involving $\beta_b$ when $d=4$. Although a
direct derivation is at present lacking by considering the results for
free fields using (29) we may obtain
$$ \beta_b = {\pi^4\over 64\times 540} ( 2\beta -3\gamma -95\ep ) \, .
\eqno (39) $$

An important issue is whether there are any positivity constraints on the
energy momentum tensor three point function. From positivity of the two
point function in unitary theories it is necessary that the scale factor
$C_T > 0$. As pointed out by Cappelli and Latorre [15] it is possible to obtain
stronger conditions by making further assumptions on the matrix elements of
the energy momentum tensor. If we define $\Theta$ as the antilinear
conjugation operator combined with reflection in the plane $x_1=0$,
$\Theta^2=1$, and for
$\Psi$ denoting some function of the fields restricted to $x_1>0$ then
the requirement of reflection positivity for Euclidean theories is just
$\l \Psi \, \Theta \Psi \r > 0$. The requirement of positive
energy density may then be translated after analytic continuation to the
Euclidean regime into
$$ \l \Psi \, T_{11}(0) \, \Theta \Psi \r < 0 \, .
\eqno (40) $$
This is of course stronger than the requirement that the Hamiltonian
operator $H = - \int \! d^d x \, \de (x_1) \, T_{11} (x)$ should have a
positive spectrum. Since $\Theta T_{i1}(y) = - T_{i1}({\bar y})$ and
$\Theta T_{ij}(y) = T_{ij}({\bar y})$ then from (27) this requires
$$ \beta > 0 \, , \quad \gamma  > 0 \, , \quad -\ep > 0 \, .
\eqno (41) $$
{}From (28) and (29) these conditions are satisfied for free theories
for $d\ge 2$ and also from (34) they imply $C_T > 0$. Unfortunately
(39) indicates that there is no consequence that $\beta_b > 0$ contrary
to known results for free theories ($64\times 90\pi^2 \beta_b =
n_S + 11 n_F + 62 n_V$). Perhaps a more complete analysis and/or
imposition of stronger conditions such as quantum analogues of the energy
conditions in classical general relativity [16] may give more information.

The above results may be relevant in discussing possible critical points
in quantum field theories. Of course in physically realistic
quantum field theories  conformal invariance is broken by renormalisation
effects arising from the need of some short distance cut off even if the
classical theory is conformally invariant. As is well known the condition
of scale invariance is transmuted into the renormalisation group or
Callan-Symanzik equation where there are anomalous dimensions and also
the $\beta$ function, reflecting the induced scale dependence of the
couplings, is necessary. Since conformal invariance is much stronger than
simple scale invariance it is interesting to discuss if there is any
corresponding implications in quantum field theories. The problem with
conformal transformations as compared with scale transformations is that, as
is easily seen in (2) and (3) for $b_\mu \ne 0$, they lead to an effective
local rescaling of the metric. However on curved space it is possible to
formulate a local renormalisation group equation if besides the metric
$g_{\mu\nu}(x)$ local $x$-dependent couplings are introduced [17]. Although
additional counterterms involving derivatives of the couplings are necessary
renormalisability is unaffected (these counterterms are straightforward to
calculate for standard theories to one and two loops and are related to
additional local pieces in renormalisation group equations for local operators
and their products [18]). For theories with a single
dimensionless coupling $g$, like QCD, then the basic local renormalisation
group equation and also the equation expressing invariance under
diffeomorphisms have the schematic forms
$$ \eqalign{
\int \! d^d x \, \si \Bigl ( 2 g^{\mu\nu} {\de \over \de g^{\mu\nu}}
- \beta (g) {\de\over \de g} \Bigl ) W = {}&0 \, , \cr
 \int \! d^d x \, \Bigl ( \L_v g^{\mu\nu} {\de \over \de g^{\mu\nu}}
+ v {\cdot \pr} g {\de\over \de g} \Bigl ) W = 0 {}&\, , \quad
\L_v g^{\mu\nu} = - \nab^\mu v^\nu -\nab^\nu v^\mu \, , \cr}
\eqno (42) $$
where $W$ is the vacuum self energy depending on the metric $g^{\mu\nu}$
and coupling $g$. Clearly for non constant $\si(x)$
consistency demands a non constant $g(x)$.
For conformal transformations $\L_v g^{\mu\nu} = -2\si_v g^{\mu\nu}$ these
equations may be combined and restricted to flat space  giving
$$ T^g{}_{\! v} W = 0 \, , \quad T^g{}_{\! v} = \int \! d^dx \,
\bigl ( - v{\cdot \pr} g + \si_v \beta(g) \bigl){\de\over \de g} \, ,
\eqno (43) $$
where now $v_\mu$ satisfies (2) and hence has the explicit form shown in (3).

Such equations may be extended to $n$-point functions for arbitrary
operators $\O_i$. For simplicity we consider only the application to two point
functions for operators which are quasi-primary in the conformal limit
when the corresponding
broken conformal invariance equation may be written as
$$ \eqalign {
\T_v \l \O_1 (x) \, \O_2 (y) \r = {}& 0 \, ,\quad
\T_v = T_{1\, v} (x) + T_{2\, v} (y) + T^g{}_{\! v} \, , \cr
T_{i\, v}(x)&{} =  v_\mu (x) {\pr \over \pr x_\mu} + \eta_i(g(x)) \si_v(x)
+ \half s_{i\, \mu\nu} \omega_{v\mu\nu} (x) \, . \cr}
\eqno (44) $$
It is easy to check that the full generator corresponding to a conformal
transformation appearing in (44) satisfies the required Lie
algebra $[\T_{v_1} , \T_{v_2} ] = \T_{[v_1 , v_2]}$. At a fixed point,
$\beta(g_*) = 0$ with $g_*$ constant, this reduces to the condition for
conformal invariance, with dimensions $\eta_i = \eta_i(g_*)$, which was
solved earlier as in (11). In general for $g$ constant and also $\si_v=0$
(44) implies rotational and translational invariance for
$\l \O_1 (x) \, \O_2 (y) \r$ while for scale transformations where
$\si_v$ is constant then the usual renormalisation group equation is
recovered
$$ \Bigl ( s {\cdot {\pr\over \pr s}} + \beta(g) {\pr\over \pr g}
+ \eta_1 (g) + \eta_2(g) \Bigl ) F(s,g) = 0 \, , \quad
\l \O_1 (x) \, \O_2 (y) \r = F(x-y,g) \, .
\eqno (45) $$
In general (44) requires $x$-dependent couplings which is not very
useful. However additional information may be obtained by considering an
expansion in derivatives of $g$ at some convenient point. For
$\O_1 , \, \O_2$ scalar fields then we may thus write, with $s=x-y$,
$$ \l \O_1 (x) \, \O_2 (y) \r = F(s,g({\bar x})) + F_1(s,g({\bar x}))
s{\cdot \pr} g({\bar x}) + \dots \, , \quad {\bar x} = \half ( x+y) \, ,
\eqno (46) $$
where $F(s,g), \, F_1(s,g)$ are just scalar functions of $s^2$. If we take
$v_\mu \to b_\mu x^2 - 2 x_\mu b{\cdot x}$ and hence $\si_v \to -2 b{\cdot x}$
then (44) gives the exact relation for terms of zeroth order in $\pr_\mu g$
$$ \eqalign {
\Bigl (& x^2 {\pr \over \pr x_\mu} - 2 x_\mu \, x {\cdot {\pr \over \pr x}}
+ x^2 {\pr \over \pr y_\mu} - 2 y_\mu \, y {\cdot {\pr \over \pr y}} \cr
& - 2 x_\mu \, \eta_1(g) - 2 y_\mu\, \eta_2(g) - (x_\mu + y_\mu )
\beta(g) {\pr\over \pr g} \Bigl ) F(s,g) = 2 s_\mu \, F_1(s,g) \, . \cr}
\eqno (47) $$
Using the dependence of $F, \, F_1$ on $s^2$ this result is equivalent
to (45) again together with the additional relation
$$ \bigl ( \eta_1(g) - \eta_2(g) \bigl ) F(s,g) = - 2 \beta(g) \,
F_1(s,g) \, .
\eqno (48) $$
Of course (48) implies that at a critical point $F(s,g)$ must vanish if
$\eta_1(g_*) \ne \eta_2(g_*)$ which coincides with the well known
consequence of conformal invariance that the two point function must be
zero if the operator dimensions are different.
However this relation also has content
away from any critical point since $F_1(s,g)$ is perturbatively calculable.
For instance knowing $F_1(s,g)$ at $\ell$ loops is sufficient to give
$F(s,g)$ at $\ell+1$ loops if $\eta_1(0) \ne \eta_2(0)$ and $\beta(g)$ is
non zero at one loop. It is also possible from (44) to derive a standard
renormalisation group equation for $F_1$ when we obtain
$$  \Bigl ( s {\cdot {\pr\over \pr s}} + \beta(g) {\pr\over \pr g}
+ \eta_1 (g) + \eta_2(g) + \beta'(g) \Bigl ) F_1 (s,g)
+\half \bigl ( \eta_1{}' (g) - \eta_2{}'(g) \bigl ) F(s,g) =  0 \, ,
\eqno (49) $$
which is easily seen to be compatible with (45) and (48). Note that for
$\O_1 = \O_2$ $F_1(s,g)=0$ which is necessary for symmetry under
$x \leftrightarrow y$.

It is also of interest to consider the case of the two point function
for a vector operator $V_\mu$. In this case we may write for local couplings
$g(x)$
$$ \eqalign {
\l V_{\mu} (x) \, V_\nu (y) \r = {}& \de_{\mu\nu}\,  F_1(s,g({\bar x}))
- 2 {s_\mu s_\nu \over s^2} \,  F_2(s,g({\bar x})) \cr
{}& + \bigl ( s_\mu\pr_\nu g({\bar x}) - s_\nu \pr_\mu g({\bar x}) \bigl )
F_3(s,g({\bar x}))  + \dots  \, , \cr}
\eqno (50) $$
consistent with symmetry under $x \leftrightarrow y, \, \mu \leftrightarrow
\nu$. $F_i(s,g)$ are scalar functions of $s^2$.
In this case the broken conformal invariance equation (44) includes
the appropriate spin generators, $(s_{\mu\nu})_{\alpha\beta} = -
\de_{\mu\alpha}\de_{\nu\beta} + \de_{\mu\beta}\de_{\nu\alpha}$.
The standard renormalisation group equations which follow from (44) are
$$ \eqalign {
\Bigl ( s {\cdot {\pr\over \pr s}} + \beta(g) {\pr\over \pr g}
+ 2\eta (g) \Bigl ) F_{1,2} (s,g) = 0 \, , \cr
\Bigl ( s {\cdot {\pr\over \pr s}} + \beta(g) {\pr\over \pr g}
+ 2\eta (g) + \beta'(g) \Bigl ) F_3 (s,g) = 0 \, . \cr}
\eqno (51) $$
In addition we may obtain, analogously to (48),
$$ 2\bigl ( F_1(s,g) - F_2(s,g) \bigl ){} = \beta (g) F_3(s,g) \, ,
\eqno (52) $$
which shows how at a critical point $F_1 = F_2$ in accord with the constraints
of conformal invariance as exemplified in the explicit form in (14) and (12).
It is interesting to consider the conditions flowing from the conservation
equation $\pr_\mu V_\mu =0$. From (50) it is easy to see that not only
$$ s {\cdot {\pr\over \pr s}} \bigl ( F_1(s,g) - 2F_2(s,g) \bigl ){} -
2(d-1) F_2(s,g) = 0
\eqno (53) $$
but also, taking account the $x$ dependence of the couplings,
$$ \eqalign {
{\pr\over \pr g} F_1(s,g) + (d-1) F_3(s,g) + s {\cdot {\pr\over \pr s}}
F_3(s,g) = 0 \, , \cr
2 {\pr\over \pr g} F_2(s,g) + s {\cdot {\pr\over \pr s}} F_3(s,g) = 0 \, . \cr}
\eqno (54) $$
If these conservation equations are combined with (52) and the renormalisation
group equations (51) it is easy to see that consistency is only possible
if $\eta (g) = d-1$, just as in pure conformal theories.

We should remark that Kraus and Sibold [19] have also considered
renormalisation
modified generators of special conformal transformations of the same form as
shown in (44) in the context of $\phi^4$ theory in four dimensions. Both
treatments are based on considering $x$ dependent couplings with a local
renormalisation group equation on curved spacetime although details are
very different. In the proceedings of the previous
Ahrenshoop Symposium [20]\footnote{*}
{Note that the results of [20] in (2.17,18) are identical in content with
(3.27a,b) of [17] when applied to $\phi^4$ theory, despite notational
complexities in both cases.}
Kraus and Sibold referred to our previous work [17] as `almost completely
meaningless'. We may hope that further extensions of the above will render
at least the almost as justified.
\vskip 6pt

I would like to thank Jos\'e Latorre for very useful discussions and
describing his results on positivity.

\noindent {\bf References}
\parskip=0cm
\parindent=20pt
\vskip 3pt

\item {[1]} J. Cardy, {\it in} `Champs, Cordes et Ph\'enom\`enes Critiques',
(E. Br\'ezin and J. Zinn-Justin eds.) North Holland, Amsterdam 1989.
\vskip 3pt
\item {[2]} A.B. Zamolodchikov, {\it JETP Lett.} {\bf 43} (1986) 43; {\it
Sov. J. Nucl. Phys.} {\bf 46} (1988) 1090;\hfill\break
A.W.W. Ludwig and J.L. Cardy, {\it Nucl. Phys.} {\bf B285} [FS19] (1987) 687;
\hfill\break
A. Cappelli and J.I. Latorre, {\it Nucl. Phys.} {\bf B340} (1990) 659.
\vskip 3pt
\item {[3]} A. Cappelli, D. Friedan and J.I. Latorre, {\it Nucl. Phys.}
{\bf B352} (1991) 616.
\vskip 3pt
\item {[4]} J. Polchinski, {\it Nucl. Phys.} {\bf B303} (1988) 226.
\vskip 3pt
\item {[5]}  K. Lang and W. R\"uhl, {\it Z.Phys.} C {\bf 50} (1991) 285;
{\bf 51} (1991) 127; {\it Nucl. Phys.} {\bf B377} (1992) 371;
{\bf B400} [FS] (1992) 597; {\bf B402} (1993) 573.
\vskip 3pt
\item {[6]} A. Petkou and H. Osborn, preprint DAMTP 93/31, hep-th/9307010,
{\it Ann. Phys. (N.Y.)}, to be published.
\vskip 3pt
\item {[7]} G. Mack and A. Salam, {\it Ann. Phys. (N.Y.)} {\bf 53} (1969) 174.
\vskip 3pt
\item {[8]} S. Ferrara, R. Gatto, A.F. Grillo and G. Parisi, {\it Lett. al
Nuovo Cim.} {\bf 4} (1972) 115.
\vskip 3pt
\item {[9]} G. Mack, {\it Comm. Math. Phys.} {\bf 53} (1977) 155.
\vskip 3pt
\item {[10]} J. Cardy, {\it Nucl. Phys.} {\bf B290} (1987) 355.
\vskip 3pt
\item {[11]} E. Schreier, {\it Phys. Rev.} {\bf D3} (1971) 980.
\vskip 3pt
\item {[12]} Ya.S. Stanev, {\it Bulg. J. Phys.} {\bf 15} (1988) 93.
\vskip 3pt
\item {[13]} D.Z. Freedman, K. Johnson and J.I. Latorre, {\it Nucl. Phys.}
{\bf B371} (1992) 353.
\vskip 3pt
\item {[14]} M. Duff, `Twenty Years of the Weyl Anomaly', preprint
CTP-TAMU-06/93,  hep-th/930807, {\it Classical and Quantum Gravity}, to
be published.
\vskip 3pt
\item {[15]} J.I. Latorre, private communication.
\vskip 3pt
\item {[16]} S.W. Hawking and G.F.R. Ellis, `The Large Scale Structure
of Space-time', Cambridge University Press, Cambridge 1973.
\vskip 3pt
\item {[17]} H. Osborn, {\it Nucl. Phys.} {\bf B363} (1991) 486.
\vskip 3pt
\item {[18]} I. Jack and H. Osborn, {\it Nucl. Phys.} {\bf B343} (1990) 647.
\vskip 3pt
\item {[19]} E. Kraus and K. Sibold, {\it Nucl. Phys.} {\bf B398} (1993) 125.
\vskip 3pt
\item {[20]} E. Kraus and K. Sibold, Proceedings of the XXVI International
Symposium Ahrenshoop, preprint DESY 93-013.

\bye